\documentclass[prl,twocolumn,showpacs]{revtex4}

\newcommand{\sub}[1]
{_\textrm{\scriptsize #1}}

\begin{document}

\title{Correlated Trapped Bosons and the Many-Body Efimov Effect}

\author{O.~S\o rensen}
\author{D.~V.~Fedorov}
\author{A.~S.~Jensen}

\affiliation{Institute of Physics and Astronomy, University of Aarhus,
  DK-8000 Aarhus C, Denmark}

\date{\today}

\begin{abstract} 
  We study two-body correlations in systems of identical bosons. We
  use a Faddeev type of decomposition of the wave function where all
  pairs of particles are treated equally.  We focus on a new
  multi-particle Efimov effect at large scattering length, where
  infinitely many loosely bound many-body states appear.  A confining
  external trap only allows a finite number of such spatially extended
  negative energy states inside the trap.  The stability of a Bose
  condensate is determined by the decay into these model independent
  intermediate states which in turn decay into dimers.
\end{abstract}


\maketitle

\paragraph*{Introduction.}

The novel theoretical formulation in \cite{sor01} was constructed to
describe correlations in boson systems and applied for a realistic
short-range repulsive interaction to Bose-Einstein condensates. The
method goes beyond the mean-field approximation \cite{dal99,bay96} and
is as well applicable to attractive finite range potentials with very
large scattering lengths where the Efimov effect occurs
\cite{efi70,fed93}.  Experimental properties are available for various
condensates from the first observations of both effectively repulsive
\cite{and95} and attractive \cite{bra95} interatomic interactions to the
recent observations \cite{cor00,rob01} of condensates in a magnetic
field used to tune the effective interaction via a Feshbach resonance
to almost any scattering length.

The condensate \cite{dal99} being an excited state of the full
many-body system is clearly unstable. The three-body recombination
into bound dimers is a dominating decay channel
\cite{nie99,esr99,bra02b}.  In the condensate this occurs independently
for neighbouring pairs and much more frequently in a coherent process
best described as a collective or macroscopic collapse
\cite{pit96,adh02}.  Two-body correlations therefore must be crucial
for this collapse, which becomes more likely and eventually inevitable
as the scattering length is increased.  Unfortunately a theoretical
description is hindered by the difficulties, especially pronounced for
large scattering lengths, of finding the decisive correlated structure
\cite{dal99,blu01,cow01}.

A promising form of a correlated wave function suggested for nucleons
\cite{rip84} was recently extended to more general systems
\cite{bar99a}.  Another related formulation uses generalized
hyperspherical coordinates and an adiabatic expansion with the
hyperradius as the adiabatic coordinate \cite{boh98}. It was applied
for many identical bosons, still only with a zero-range interaction
and the lowest (constant and thus non-correlated) hyperspherical
angular wave function. These crude approximations are removed in a
novel method still using hyperspherical coordinates and adiabatic
expansion, but now the wave function is determined from a
variationally established equation \cite{sor01}.  The purpose of this
letter is to investigate the structure of boson systems as a function
of (large) scattering length for attractive finite range potentials
with emphasis on an emerging novel many-body Efimov effect.

\paragraph*{Theory.}

A system of $N$ identical particles of masses $m$, trapped in an
external field approximated by a harmonic oscillator potential of
angular frequency $\omega$, can in the center of mass frame be
described by hyperspherical coordinates, i.e.  $3N-4$ hyperangles
\cite{bar99a,bar99b} denoted collectively by $\Omega$ and one
hyperradius, $\rho$, given by \cite{sor01}
\begin{eqnarray}
  \label{eq:0}
  \rho^2&=&\frac1N\sum_{i<j=1}^Nr_{ij}^2=\sum_{i=1}^Nr_i^2-NR^2
  \;,
\end{eqnarray}
where $\vec r_i$ are single-particle coordinates, $\vec R$ is the
center of mass coordinate, and $r_{ij}=|\vec r_j-\vec r_i| \equiv
\sqrt2\rho\sin\alpha_{ij}$ with $\alpha_{ij}$ varying between $0$ and
$\pi/2$. The hamiltonian is
\begin{eqnarray}
  \label{eq:2}
  \hat H&=&
  \sum_{i=1}^N\Big(\frac{\hat p_i^2}{2m}+\frac12m\omega^2r_i^2\Big)+
  \sum_{i<j=1}^NV(r_{ij}) \; ,
\end{eqnarray}
where $V$ is the two-body interaction. It separates into a center of
mass part, a radial part, and an angular part depending respectively on
$\vec R$, $\rho$, and $\Omega$ \cite{sor01}:
\begin{eqnarray}
  \label{eq:2b}
  \hat H =
  \hat H\sub{cm}
  +
  \hat H_\rho+\frac{\hbar^2\hat h_\Omega}{2m\rho^2}
  \;,
\end{eqnarray}
with $\hat H\sub{cm}=\hat p_R^2/(2Nm)+Nm\omega^2R^2/2$, $\hat
H_\rho=\hat T_\rho+m\omega^2\rho^2/2$ and $\hbar^2\hat
h_\Omega/(2m\rho^2)=\hat T_\Omega+\sum_{i<j}V_{ij}$, where $\hat
T_{\rho}$ and $\hat T_{\Omega}$ are related kinetic energy operators.
A suitable expansion of the wave function is
\begin{eqnarray}
  \label{eq:3}
  \Psi&=&\rho^{-(3N-4)/2}\sum_{n=0}^\infty
  f_n(\rho)\Phi_n(\rho,\Omega)
  \;,
\end{eqnarray}
where $\Phi_n$ is an eigenfunction of the angular part of the
hamiltonian with an eigenvalue $\hbar^2\lambda_n(\rho)/(2m\rho^2)$:
\begin{eqnarray}
  \label{eq:4}
  \hat h_\Omega\Phi_n(\rho,\Omega)&=&\lambda_n(\rho)\Phi_n(\rho,\Omega)
  \;.
\end{eqnarray}
Neglecting couplings between the different $n$-channels yields the
radial eigenvalue equation for the energy $E_n$:
\begin{eqnarray}
  \label{eq:5}
  \Big(-\frac{\hbar^2}{2m}\frac{d^2}{d\rho^2} + U_n(\rho) - E_n\Big)
  f_n(\rho)   &=& 0
  \;,
  \\
  \label{eq:6}
  U_n(\rho)=\frac{m\omega^2\rho^2}2 +
  \frac{\hbar^2(3N-4)(3N-6)}{8m\rho^2} &+&
  \frac{\hbar^2\lambda_n}{2m\rho^2}
  \;.\quad
\end{eqnarray}
The second term in the radial potential $U$ is a generalized
centrifugal barrier. We now decompose the angular wave function $\Phi$
in the symmetric Faddeev components $\phi$
\begin{eqnarray}
  \label{eq:7}
 \Phi(\rho,\Omega) = \sum_{i<j=1}^N  \phi_{ij}(\rho,\Omega) \approx
 \sum_{i<j=1}^N  \phi(\rho,r_{ij})
  \;,
\end{eqnarray}
where the restricting assumption is that the interparticle potentials
only act in $s$-waves leaving only the dependence on the distance
$r_{ij} = \sqrt2\rho\sin\alpha_{ij}$.  The capability of this
decomposition for large scattering length has been demonstrated for
$N=3$ by the proper description of the intricate Efimov effect
\cite{fed93,jen97}.

The eigenvalue equation eq.~(\ref{eq:4}) can by a variational
technique be rewritten as a second order integro-differential equation
in the variable $\alpha_{12}$ \cite{sor01}.  For atomic condensates the
interaction range is very short compared to the spatial extension of
the $N$-body system. Then this equation simplifies even further to
contain at most one-dimensional integrals. The principal interaction
dependence is through the parameter $a\sub B\equiv m\int_0^\infty
drr^2V(r)/\hbar^2$, which is the Born-approximation to the $s$-wave
scattering length $a\sub s$.  The validity of our approximations only
relies on a small \emph{range} of the potential whereas the scattering
length can be as large as desired.

\paragraph*{Angular potentials.}

The crucial ingredient is the angular eigenvalue $\lambda$ obtained
from eq.~(\ref{eq:4}). We choose gaussian two-body potentials
$V(r)=V_0\exp(-r^2/b^2)$ of range $b$ and strength $V_0$.  We measure
lengths in units of $b$ and energies in units of $\hbar^2/(2mb^2)$.
We solve the equation by choosing discrete sets of mesh points and
thus constructing a matrix equation ready for diagonalization
\cite{sor01}.  This computational scheme increases strongly with $N$
and we therefore illustrate by relatively small values of $N$.

\begin{figure}[htb]
  \begin{center}
    \input{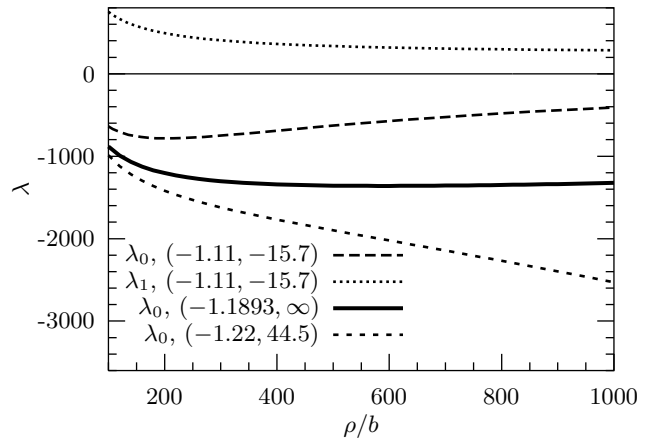} 
  \end{center}
  \vspace{-0.3cm}
  \caption [Angular eigenvalue close to two-particle threshold] 
  { Angular eigenvalues as functions of $\rho$ for $N=20$ and
    various potential strengths given by $(a\sub B/b,a\sub s/b)$.
    The long-dashed and the dotted curves correspond to the two
    lowest eigenvalues when the two-body potential is too weak to
    support a bound state. The potential giving rise to the
    short-dashed curve has only one bound two-body state and the
    solid curve corresponds to the two-body threshold of zero
    energy.} 
  \label{fig:lambdavar}
  \vspace{-0.3cm}
\end{figure}

Fig.~\ref{fig:lambdavar} shows the angular eigenvalues for different
strengths expressed in terms of $a\sub B$ and $a\sub s$.  For short range
potentials finite at the origin all $\lambda$ approach their values at
$\rho=0$ as $\rho^2$. This is a model dependent region and therefore
not considered in the following.  When the potential is too weak to
support a two-body bound state, ($a\sub B/b>-1.18934$), the lowest
$\lambda_0$ is negative and approaches zero as $a\sub s/\rho$ with
increasing $\rho$. In fig.~\ref{fig:lambdavar} we also show the
eigenvalue $\lambda_1$ corresponding to the first excited angular
state. It is positive and approaches the value $4(4+3N-5)$ as
$-a\sub s/\rho$ for large $\rho$ \cite{jen97,boh98}.

Increasing the attraction of the potential a two-body bound state of
energy $B_2$ appears (positive $a\sub s$) and the corresponding lowest
angular eigenvalue diverges proportional to $2 m B_2\rho^2 /\hbar^2$
when $\rho \rightarrow \infty$.  The many-body system would obviously
then prefer to recombine into bound dimer states. The first excited
angular eigenvalue, similar to the dotted curve in
fig.~\ref{fig:lambdavar}, is then responsible for the unstable state
referred to as the condensate.

At the two-body threshold the scattering length is infinite. Then the
lowest $\lambda$ approaches a negative constant $\lambda_\infty$ for
large $\rho$ ($\rho \gtrsim \rho_b \equiv 45 \sqrt Nb$), see
fig.~\ref{fig:lambdavar}. Using $N=3,10,20$ we find numerically
$\lambda_\infty(N) \approx 24 (N-2) - 5 N(N-1)$, where the form
reflects our expectations of $N$-dependence of kinetic and potential
energies.

These angular eigenvalues behave qualitatively different from those
obtained in \cite{boh98}, where the expectation value of a
$\delta$-interaction of strength proportional to $a\sub s$ is computed
with a constant angular wave function without any correlations. The
eigenvalues are then necessarily proportional to $a\sub s/\rho$ for
all $\rho$, i.e. always diverging when $\rho \rightarrow 0$ and
converging to zero for $\rho \rightarrow \infty$.  The differences are
especially pronounced for large scattering lengths and when a bound
two-body state is present.  However, even in the qualitatively similar
case in fig.~\ref{fig:lambdavar} of $a\sub s/b = -15.7$ our $\lambda$ is
lower than that of \cite{boh98} by a factor between $1.5$ and $2$ for
$\rho/b\in[10^3,10^5]$.

At large $\rho$, $\Phi$ approaches a non-correlated angular wave
function.  Using a constant as in \cite{boh98} the expectation value
of our gaussian interaction gives for large $\rho$ and $N$ that
$\lambda = \lambda\sub{n-c}(\rho) \equiv kN^{7/2}a\sub B/\rho$ with $k
= (3/2)^{3/2}/\sqrt{2\pi}$, i.e. we get $a\sub B$ instead of $a\sub s$ as
assumed in \cite{boh98}.  For large $a\sub s$ our computed $\lambda$ bends
upwards from the plateau of $\lambda_\infty$ around a point $\rho
\equiv \rho\sub{th}$ determined by the large but finite value of $a\sub s$.
If $\rho > \rho\sub{th}$ then $\lambda$ can be estimated as
$\lambda\sub{n-c}(\rho)$ with $a\sub s$ substituted for $a\sub B$. The
transition point is then given by $\lambda_\infty =
\lambda\sub{n-c}(\rho\sub{th})$ which for large $N$ yields
\begin{eqnarray}
  \label{eq:9} 
 \rho\sub{th}=
  \frac{1}{5} k|a\sub B| N^{3/2}
  \simeq
  \frac17|a\sub B|N^{3/2} \rightarrow   \frac17|a\sub s|N^{3/2}
  \;,
\end{eqnarray}
where the more realistic estimate for attractive potentials is
indicated by exchanging $a\sub B \rightarrow a\sub s$, see
\cite{jen97,boh98}.

\paragraph*{Radial potentials.}

In eq.~(\ref{eq:6}) the external harmonic trap of angular frequency
$\omega$ corresponds to a trap length $b\sub
t\equiv\sqrt{\hbar/(m\omega)}$.  We model the experimentally studied
systems \cite{and95,rob01} of $^{85}$Rb and $^{87}$Rb-atoms with
oscillator frequencies $\nu=\omega/(2\pi)=205$ Hz and $200$ Hz and
interaction range $b=10$ a.u., thus yielding $b\sub t/b=1442$.

\begin{figure}[htb]
  \begin{center}
    \input{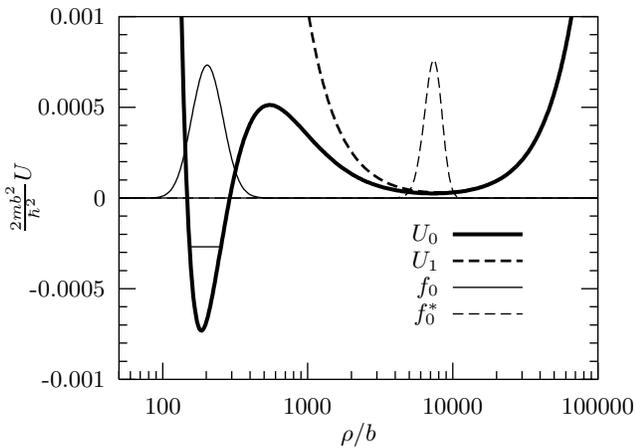}
  \end{center}
  \vspace{-0.3cm}
  \caption [Angular eigenvalue close to two-particle threshold]
  {Radial potentials $U_0$ and $U_1$ from eq.~(\ref{eq:6})
    corresponding to the two lowest angular potentials for $N=20$
    and $a\sub s/b = -15.7$ in fig.~\ref{fig:lambdavar}. The trap
    length is $b\sub t/b=1442$. Also shown are the two lowest
    energies and the radial eigenfunctions $f_0$ and $f_0^*$ for the
    lowest radial potential.}
  \label{fig:uasneg} 
  \vspace{-0.3cm}
\end{figure}

We show in fig.~\ref{fig:uasneg} the radial potentials for the angular
eigenvalues corresponding to unbound two-body systems ($a\sub s/b = -15.7$
in fig.~\ref{fig:lambdavar}).  The external field in eq.~(\ref{eq:6})
is negligible for small $\rho$ and therefore the radial potential is
negative when $\lambda+(3N-4)(3N-6)/4<0$. Then genuinely bound states
of negative energy are possible in our model without the confinement
from the trap.

The radial potential for the lowest angular eigenvalue has a global
minimum with $U\sub{min}<0$ at small $\rho$ ($\approx \rho_b $)
separated by a barrier at intermediate $\rho$ from a local minimum at
$\rho \sim \rho\sub t\equiv\sqrt{3N/2}\;b\sub t$.  At large $\rho$ the
radial potential diverges due to the harmonic term, at small $\rho$
due to the centrifugal term.  The radial potential corresponding to
the first excited angular state, also in fig.~\ref{fig:uasneg},
coincides with the lowest radial potential at large $\rho$, but does
not contain the attraction at small $\rho$ (see
fig.~\ref{fig:lambdavar}) and diverges therefore to $+$ infinity for
small $\rho$ without going through another minimum.

The radial equation in eq.~(\ref{eq:5}) has only one solution with
negative energy $E$ and the corresponding wave function, shown in
fig.~\ref{fig:uasneg}, is located in the global minimum.  The first of
the infinitely many excited states in this potential is located in the
local minimum at larger $\rho$ created by the competition between the
centrifugal barrier and the external harmonic oscillator potential.
This excited state is usually referred to as the condensate, but the
moderate attraction for the relatively small $a\sub s = -15.7 b$ already
results in a lower lying many-body state about $\rho\sub t/\rho_b
\approx 37$ times smaller than the condensate.

\paragraph*{Large scattering length.}

At the threshold for binding of the two-body system the angular
eigenvalue outside the interaction range must approach a negative
constant $\lambda \rightarrow \lambda_\infty$, see
fig.~\ref{fig:lambdavar}.  The scattering length is then infinitely
large and the radial potential has the form \cite{nie01}
\begin{eqnarray}
  \label{eq:10}
  U(\rho)
  &\to&
  \frac{\hbar^2}{2m}\bigg(\frac{-\xi^2-1/4}{\rho^2}+
  \frac{\rho^2}{b\sub t^4}\bigg)\;,
\end{eqnarray}
where $\xi^2\equiv-\lambda_\infty-(3N-4)(3N-6)/4-1/4 \approx 11N^2/4$
for large $N$, where $\lambda_\infty \approx - 5 N^2$. Without the
external $\rho^2$ potential in eq.~(\ref{eq:10}) the infinitely many
radial solutions to eq.~(\ref{eq:6}) all behave like
\begin{eqnarray}
  f_\infty(\rho) =
  \sqrt\rho\sin\big(|\xi|\ln(\rho/\rho\sub{sc})\big)
  \;,
\end{eqnarray}
with some hyperradius scale $\rho\sub{sc}$. The energies and sizes of
the eigenstates labeled $j$ are related by \cite{efi70,nie01}
\begin{eqnarray}
  \frac{E_j}{E_{j+1}}
  &=&
  \frac{\langle\rho^2\rangle_{j+1}}{\langle\rho^2\rangle_j}
  =
  e^{2\pi/|\xi|}
  \;.
  \label{eq:1}
\end{eqnarray}
With increasing quantum number these states become exponentially
larger with exponentially smaller energies approaching zero.  They
originate from the constant $\lambda$ and the generic $1/\rho^2$
potential in eq.~(\ref{eq:10}). They are many-body states, but not the
embedded three-body Efimov cluster states which also arise from a
$1/\rho^2$ potential. However, both multi- and three-body Efimov
states appear for very large two-body scattering lengths.

\begin{figure}[htb]
  \centering 
  \input{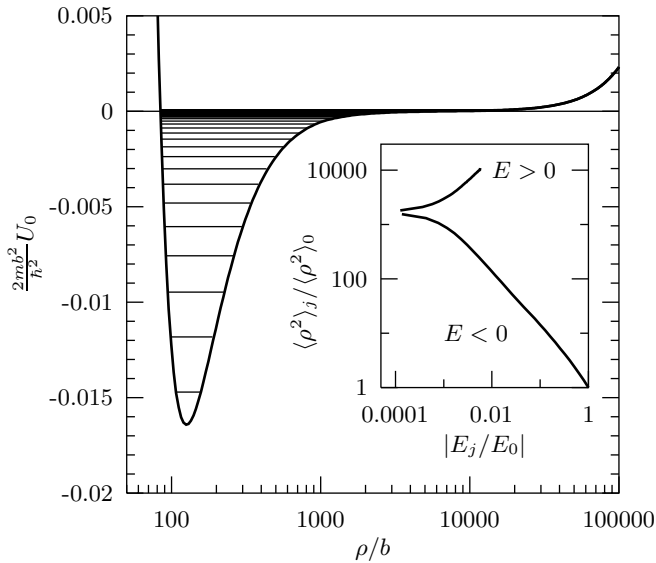}
  \vspace{-0.7cm}
  \caption
  [Threshold to bind $N$-particle system and radial properties of
  bound many-body states] {The lowest radial potential for $N=20$,
    $a\sub s/b = \infty$, and $b\sub t/b=1442$.  The horizontal lines
    indicate the $69$ (30 below zero) lowest energy eigenvalues and
    the inset shows their mean square hyperradii as a function of the
    absolute value of their energies relative to the values for the
    lowest state, i.e.  $2mb^2E_0/\hbar^2 \simeq -0.0147$,
    $\sqrt{\langle\rho^2\rangle_0}/b \simeq 136$.}
  \label{figs:asinfty} 
  \vspace{-0.3cm}
\end{figure}

The external harmonic oscillator limits the possible number of these
new states with $E<0$.  These negative energy states have to be
located inside the trap and outside the two-body potential. The number
of nodes of $f_\infty$ allowed in this region equals the number of
bound states $N_E \approx
|\xi|\pi^{-1}\ln(\rho\sub{max}/\rho\sub{min})$, where the first and
last zero points then are at the end points of the interval. We obtain
$\rho\sub{max}/\rho\sub{min} \approx \rho\sub t/\rho_b$ (see also
fig.~\ref{figs:asinfty})
\begin{eqnarray}
  \label{eq:13} 
  N_E \approx  \frac{|\xi|}\pi\ln\bigg(\frac{ b\sub t} {37 b} \bigg) 
  \;,\quad
  \textrm{for }b\sub t\ll N|a\sub s|
  \;,
\end{eqnarray}  
where the quoted condition of validity is found from $\rho\sub t \ll
\rho\sub{th}$, i.e. the plateau value must extend beyond the trap.  The
number of these available Efimov states therefore scales proportional
to $N$ ($\xi \propto N$) and logarithmically with the ratio of trap
length and interaction range.

We show in fig.~\ref{figs:asinfty} the radial potential for $N=20$ and
infinite scattering length corresponding to $\lambda_\infty\sim-1340$
or $\xi^2\sim 584$, see fig.~\ref{fig:lambdavar}. Using the estimate
in eq.~(\ref{eq:13}) we get $N_E \approx 28$ in good agreement with
the computed 30.  The lower curve in the inset of
fig.~\ref{figs:asinfty} is according to eq.~(\ref{eq:1}) a straight
line of slope $-1$.  The very lowest states deviate, since they
``feel'' non-constant $\lambda$ at small $\rho$, and the states close
to $E=0$ deviate due to the influence of the external potential.  The
energy spectrum becomes even denser above zero energy. For large
positive energies in the upper part of the inset the harmonic
potential dominates and a straight line with slope $+1$ is
obtained. The small positive energies are influenced by both external
trap and interaction potentials.

Using eq.~(\ref{eq:0}) we get $2 \langle\rho^2\rangle = (N-1) \langle
r_{12}^2\rangle \approx 2 (N-1) \langle r_{i}^2\rangle$.  Even the
most bound state, $\langle\rho^2\rangle^{1/2} \approx 136 b$, has then
a root mean square distance between two particles, $\langle
r_{12}^2\rangle^{1/2} \approx 44 b$, much larger than the interaction
range. Also the root mean square radius $\langle r_{i}^2\rangle^{1/2}
\approx 31 b$ is large.

At the plateau, $\lambda\sim\lambda_\infty$, $\xi^2>0$ and the radial
potential has no intermediate barrier. The centrifugal barrier can
only be at $\rho>\rho\sub{th}$, i.e.~in the region where
$\lambda\sim\lambda\sub{n-c}$, which provides the estimate of the
barrier position $\rho\sim 2kN^{3/2}|a\sub{s}|/3 > \rho\sub{th}$, see
eq.~(\ref{eq:9}).  A criterion of stability is then obtained by using
$\lambda\sub{n-c}$ to estimate when the corresponding radial barrier
disappears. This yields in analogy to \cite{boh98} that the condensate
will be (meta-)~stable if $N|a\sub s|/b\sub t < 0.671$.  This critical
value is from the Gross-Pitaevskii equation found to be about $0.6$
\cite{dal99,rup95}, and recently measured to $0.46$ \cite{rob01} for a
$^{85}$Rb-condensate.

Stability is determined by decay through a barrier, like in
fig.~\ref{fig:uasneg}, when negative energy states and the condensate
state in the second minimum are present.  The tunneling process then
populates the states in the first minimum, i.e. many-body states of
smaller extension and correspondingly larger density than the initial
condensate.  Competing with this macroscopic (all particles together)
tunneling are the two- and three-particle recombination processes,
which can occur from the initial condensate, or more likely from the
collapsed many-body states at larger density and smaller $\rho$. Thus
we have established a decay mechanism, i.e. macroscopic tunneling
followed by recombination processes from the intermediate model
independent states.

An even more dramatic collapse could be initiated experimentally by
first creating a condensate corresponding to fig.~\ref{fig:uasneg},
and then suddenly increasing the scattering length to that of
fig.~\ref{figs:asinfty} \cite{rob01,adh02,don01}. The multi-particle
Efimov states of fig.~\ref{figs:asinfty} would then quickly be
populated, since the barrier is removed, and these states would
subsequently leak out of the trap due to recombination.

\paragraph*{Conclusions.}

We use a new formulation designed to investigate two-body correlations
in boson systems.  The qualitative properties of Bose-Einstein
condensates are explained by using the hyperspherical adiabatic
expansion. Finite range and attractive potentials introduce new
features compared to the mean-field approximation.  Large scattering
lengths are specifically treated. Many-body Efimov states appear for
scattering lengths of the order of the trap length as
model-independent states of much smaller extension than the
condensate.  Recombination through these intermediate states is
suggested.  This opens the possibility for realistic calculations of
few-body recombination processes and for macroscopic collapse
processes.

\end{document}